\documentclass{aa}

\usepackage{graphicx}
\usepackage[varg]{txfonts}

\usepackage{natbib}
\bibpunct{(}{)}{;}{a}{}{,}
\newcommand{\Msun}{M$_{\odot}$}

\newcommand{\kms}{km\,s$^{-1}$}

\newcommand{\nii}{[N\,{\sc ii}]}
\newcommand{\oiii}{[O\,{\sc iii}]}

\newcommand{\Teff}{T_{\rm eff}}
\newcommand{\Lsun}{L$_{\odot}$}

\newcommand{\g}{\textit{Gaia}}

\defcitealias{SBJ.18}{SBJ}

\def\changed{}
\begin{document}

\title{Confronting expansion distances of planetary nebulae with \g~DR2 measurements}

\titlerunning{Expansion distances of planetary nebulae versus \g\ DR2 measurements}

\author{D. Sch\"onberner \and M. Steffen} 

\authorrunning{D. Sch\"onberner \& M. Steffen}
        
\institute{Leibniz-Institut f\"ur Astrophysik Potsdam (AIP),           
           An der Sternwarte 16, 14482 Potsdam, Germany\\
           \email{deschoenberner@aip.de}, \email{msteffen@aip.de} \label{inst1} }
           
\date{Received \today\  /  Accepted \today,  Vers. 1.4, final version}  

\abstract
     {Individual distances to planetary nebulae are of the utmost relevance for our 
      understanding of {post-asymptotic giant-branch} evolution because they allow a
      precise determination of stellar and nebular properties.   
      Also, objects with individual distances serve as  calibrators for the so-called 
      statistical distances based on secondary nebular properties.
      }        
     {With independently known distances, it is possible to check empirically 
      our understanding of the formation and evolution of planetary nebulae as suggested
      by existing hydrodynamical simulations.  
      }      
     {We compared the expansion parallaxes that have recently been determined for a number of 
     planetary nebulae with the trigonometric parallaxes provided by the \g\ Data Release 2.
     }
     {Except for two out of 11 nebulae, we found good agreement between the expansion and the
     \g\ trigonometric parallaxes without any systematic trend with distance.
     Therefore, the \g\ measurements also prove that the correction factors necessary
     to convert proper motions of shocks into Doppler velocities cannot be ignored.
     Rather, the size of these correction factors and their evolution with time as predicted by 
     1-D hydrodynamical models of planetary nebulae is basically validated.   
     These correction factors are \changed{generally} greater than unity and are different for the 
     outer shell and the inner bright rim of a planetary nebula.       
     The \g\ measurements also confirm earlier findings that spectroscopic methods often lead 
     to an overestimation of the distance.  {\rm They also show that even modelling of the 
     entire system {of star and nebula} by means of
      sophisticated photoionisation modelling may not \changed{always} provide reliable results. }
     }
     {The \g\ measurements confirm the basic correctness of the present radiation-hydrodynamics
      models, which predict that both the shell and the rim of a planetary nebula are two
      independently expanding entities, created and driven by different physical processes, 
      namely thermal pressure (shell) or wind interaction (rim), both of which vary differently
      with time.
      }
     
     
\maketitle

\section{Introduction}
\label{sec:intro}

   Individual distances to planetary nebulae (PNe) are of great relevance for our understanding
   of {post-asymptotic giant-branch} (post-AGB) evolution provided they are of 
   sufficient accuracy to allow a trustworthy 
   determination of stellar and nebular properties that can be compared with theoretical
   predictions.  Moreover, objects with known individual distances serve as calibrators for the 
   so-called statistical distances based on secondary nebular properties.  
   Prior to the \g\ era, direct trigonometric
   distances were only available for a rather limited number of close-by PNe through 
   long-term measurements of the US Naval Observatory (USNO, \citealt{harris.07}) and the 
   Hubble Space Telescope (HST, \citealt{benetal.09}).
   
   Much effort was thus invested in getting individual distances for more distant objects
   using other methods, for example detailed spectroscopic determinations of the central-star
   parameters by {non-local thermal equilibrium} 
   model-atmosphere techniques and their comparison with evolutionary tracks in the 
   (distant-independent) $\log g$-$\Teff$ plane \citep[see e.g.][and references therein]{MKH.92}.
   
   The so-called ``gravity distances`` of \citet{MKH.92} and \citet{KUP.06}
   are based on model atmospheres of different degrees of sophistication:  static atmospheres
   with limited consideration of line blanketing in the former, and  
   ``unified'' model atmospheres that also include the wind envelope in the latter work. 
   To get the distance, the stellar gravity is combined with the stellar mass that is read off 
   from post-AGB tracks in a $ \Teff / \log g $ plane, the visual absolute brightness, and the
   model flux for the given $ \Teff $ (see Eq.~4 in \citealt{MKH.92}).  

   The distances of \citet{MKH.92} and \citet{KUP.06} are based on
   the old post-AGB evolutionary tracks of \citet{schoen.79, schoen.81, schoen.83}. 
   The new evolutionary calculations of \citet{miller.16} give 
   somewhat higher post-AGB luminosities (and lower gravities) for a given remnant mass.  
   In short, the gravity distances are now smaller by about 5\%, 
   and these adjusted  distances are used here for comparison.\footnote
   {For more details, the reader is referred to Sect.~5.2 in \citet{SBJ.18}.}

   \citet{PHM.04} analysed the UV spectra of a number of PN central stars using a very
   sophisticated, hydrodynamically consistent model, which includes the expanding stellar 
   atmosphere with the supersonic wind region.  
   This model, originally developed for mass-losing massive stars,
   provides the stellar parameters (effective temperature, radius, luminosity, mass) 
   independently of any stellar evolutionary tracks, and hence also the distance.
   
   Another important approach to get an individual distance is to model the whole system, 
   central star and nebular envelope, by employing for instance the well-known photoionisation
    code ``Cloudy''\footnote
    {See www.nublado.org.}.  
   Since many parameters determine the nebular emission, a consistent
   solution that also includes  the stellar parameters is rather complex and prone to degeneracy,
   which must be resolved by additional constraints.
   An interesting variant is the combination of a full spectroscopic analysis of the stellar 
   atmosphere including the stellar wind with a consistent nebular photoionisation model, as
   has been performed for instance by \citet{MG.09} for the IC~418 system.   
   These authors also provide an illuminating discussion of the degeneracy problem and a possible
   way to solve it.    
     
   A further powerful method of deriving individual distances to planetary nebulae is the 
   expansion-parallax method.  In the latest study of this kind \citep[][hereafter SBJ]{SBJ.18},
   two- or three-epoch HST images were employed.  From these images, angular proper motions of 
   for instance the nebular rim and shell edges were determined and combined with measured
   expansion (Doppler) velocities to derive directly the distances by assuming that the 
   expansions along the line of sight and in the plane of sky are equal.   
   
   It turned out that the (corrected) expansion distances as derived by \citetalias{SBJ.18} are 
   in general smaller than the distances based on the spectroscopic methods.        
   However, both the gravity and the expansion distances are subject to
   uncertainties that are difficult to control.  In the case of the former, the
   distances rely on a precise determination of the stellar gravity 
   \citep[][Eq.~4 therein]{MKH.92}, which is a difficult task for very hot stars and the 
   main source for the distance uncertainties.
   
   The situation is even more complicated for the expansion parallaxes because we are dealing in
   this case with expanding gaseous shells led by shocks.  
   The problem, usually ignored in the past, 
   lies in the conversion of pattern expansions (the leading shocks) into flow velocities 
   (Doppler line split).  The correction factor (see Sect.~\ref{sec:ess.exp})
   to be applied for the distance is always larger than unity, but its size
   fully relies on a proper description of the formation and expansion of PNe and their 
   shock fronts.    Although it has been shown by \citet{schoenetal.14} (see also 
   \citealt{schoen.16}) that their \hbox{1-D} models give an {astonishingly} good 
   description of the nebular expansion properties, the question of whether they are also good 
   enough for distance determinations remains still to be tested, especially if one considers
   that part of the investigated PNe have a shape that is quite far from being spherical.

   With the launch of the \g\ satellite \citep{gaia.16a} it became possible to
   enlarge considerably the number of planetary nebulae with accurate trigonometric parallaxes 
    \citep{gaia.16, gaia.18}.     
   Already \g\ Data Release 1 \citep[DR1;][]{gaia.16} contains distances to PNe, albeit
   with still rather large errors for objects above 1~kpc \citep[][Fig.~1 therein]{stangh.17}.  
   The situation improved considerably with 
   \g\ Data Release 2 (DR2), and \citet{KB.18} compared the \g\ parallaxes with ground-based
   (USNO) and space-based (HST) trigonometric parallaxes and with the statistical 
   distance scales of \citet{StH.10} and \citet{frew.16}.   The agreements are very satisfying:
\begin{itemize}
\item  The USNO parallaxes are confirmed by the \g\ DR2, 
       although the USNO errors are comparatively high for objects further away than 0.5~kpc.  
       The HST distances slowly deviate with increasing distance from the 1:1 relation 
       for unknown reasons, until they are about 30\% higher at 0.50~pc 
       (cf. Fig.~1 in \citealt{KB.18}).
\item  The comparison with the two statistical distance scales reveals only insignificant
       deviations from the 1:1 relation -- though the individual distance differences can be 
       very high, up to about a factor of two to either side 
       (cf. Figs.~2 and 3 in \citealt{KB.18}).       
\end{itemize}   

    The purpose of the present paper is a comparison of the most recent expansion distances 
    determined by \citetalias{SBJ.18} with the trigonometric distances provided by the \g\ DR2.    
    As an aside, we also discuss briefly the quality of spectroscopic (or gravity) methods and/or 
    the use of photoionisation models for distance determinations for the objects in common.
      
   The paper is organised as follows.  Firstly, we present a brief introduction to the method
   of determining expansion distances (Sect.~\ref{sec:ess.exp}).
   Then, in Sect.~\ref{sec:sample.comp}, we introduce our 
   sample of PNe that have expansion and \g\ trigonometric parallaxes and compare the
   distances in detail. In particular, we demonstrate the importance of the correction
   factors.  The following Sect.~\ref{sec:empir.corr} deals with an empirical determination of
   individual correction factors, both for nebular shell and rim.
   This article closes with a short summary and the conclusion (Sect.~\ref{sec:sum.con}).

\section{Essence of the expansion method}
\label{sec:ess.exp}

\begin{figure}
\center
\includegraphics[trim= 0cm 2cm 0cm 2cm, width= \linewidth, clip]{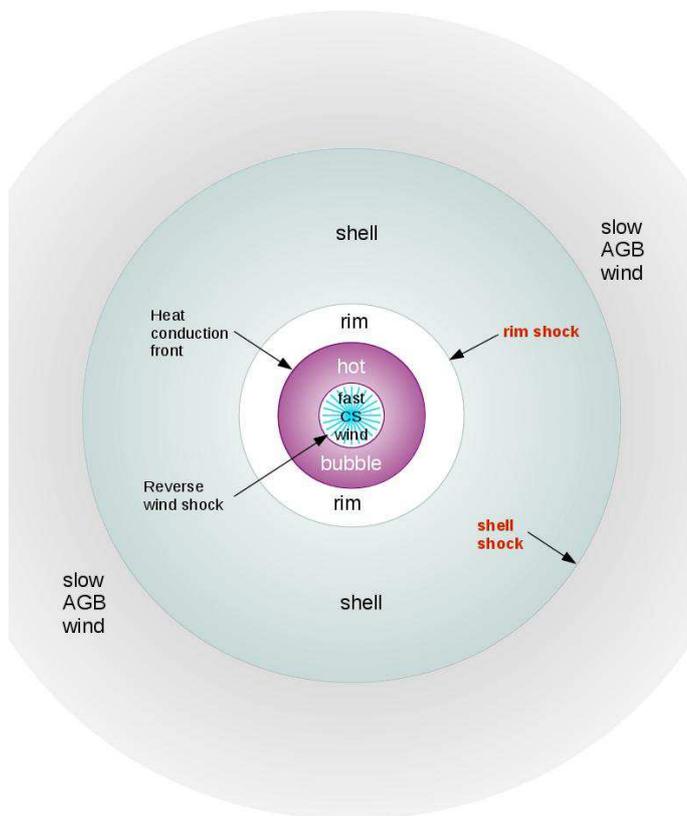}
\caption{\label{fig:sketch}      
        Sketch of the main constituents of a PN mentioned in the text; it is not to scale.
        }
\end{figure}

   We follow here the notation used in \citetalias{SBJ.18}.
   Approximating the main structure of a PN as a system of (spherical) expanding shock waves, 
   the so-called ``shell'' followed by the ``rim'' with internal velocity and 
   density gradients (cf. Fig.\ref{fig:sketch}), the distance is usually determined by the relation 
\begin{equation}
\label{eq:Dexp}
   D_{\rm exp}^0 = 211\, V_{\rm Doppler}\,/\,\dot{\theta} ,
\end{equation}
   where $ D_{\rm exp}^0 $ is the distance in parsec (pc), $ V_{\rm Doppler} $ (in \kms)  is half 
   the Doppler split of a suitable emission line observed in the direction to the nebular centre, 
   $ \theta $ (in milli-arcseconds, or mas) the angle between the centre of the nebula 
   and the feature (or nebular edge) being tracked (usually along the semi-minor axis), and
   \changed{$\dot{\theta}$ is the angular expansion rate (in mas/yr) of this feature.} 
   However, we already emphasised in \citetalias{SBJ.18} 
   that Eq.~(\ref{eq:Dexp}) is physically not correct because 
   (i) pattern velocities, for example shock fronts, are compared with matter (Doppler)
       velocities, and 
   (ii) observed Doppler splits yield density and projection weighted velocities only.  

   Instead, the physically correct version of Eq.~(\ref{eq:Dexp}) is 
\begin{equation} \label{eq:palen.corr}
   D_{\rm exp} = 211\, \dot{R}_{\rm shell/rim} /\dot{\theta}_{\rm shell/rim},
\end{equation}  
   where $\dot{R}_{\rm shell/rim}$ are the true expansion velocities of the rim or shell edges 
   (in \kms) and $\dot{\theta}_{\rm shell/rim}$ the corresponding angular expansions in the
   plane of sky (in mas\,yr$^{-1}$).  The true edge (shock) expansion velocities cannot be 
   measured and must be gained via the measured line-of-sight Doppler velocities and appropriate 
   correction factors:
   $ F =  \dot{R}_{\rm shell/rim} / V_{\rm shell/rim} > 1$, so that  
\begin{equation}
\label{eq.Dexp.corr}
   D_{\rm exp} = F_{\rm shell/rim}\times 211\, V_{\rm shell/rim} / \dot{\theta}_{\rm shell/rim} 
               = F_{\rm shell/rim}\times D_{\rm exp}^0\, .
\end{equation}     
   One must distinguish between shell and rim because it is a priori not guaranteed that the same
   correction factor is valid for shell and rim.   In any case, shell and rim should give 
   the same expansion distance if the underlying physical model is correct.

   The need for these correction factors is usually ignored in the literature, although their
   significance has already been pointed out by \citet{MGS.93}. Later these corrections were
   quantified analytically by \citet{mellema.04} and by means of hydrodynamical simulations by
   \citet{schoenetal.05} and \citetalias{SBJ.18}. 
   They are important because they increase the distance considerably, namely 
   from about 30\,\% up to about 100\,\%, depending on the expansion state of the object 
   considered, namely the properties of the respective shocks.     
  
   Detailed and internally consistent 
   1-D radiation-hydro\-dyn\-amics simulations of planetary-nebula formation and evolution 
   \citep[][]{schoenetal.97, perinetal.98, perinetal.04, VMG.02, SS.06} suggest that indeed
    the physical system of a planetary nebula mainly consists of two 
    nested shock waves that expand independently from each other.  
    The faint (outer) shell, which also contains most of the nebular mass, is
    driven by thermal pressure while the optically bright inner rim is the result of wind 
    interaction: a bubble of hot, shocked-wind gas expands against the thermal pressure of the
    ionised shell gas. A sketch of the main constituents of a PN is shown in Fig.~\ref{fig:sketch}.
    
    In this paradigm it is important to realise that the shell's shock is completely
    decoupled from the stellar wind by the rim.   The stellar wind, with possible asymmetries
    introduced by a binary central star already earlier on the AGB, only sets the stage and 
    influences, but does not control, the further dynamical evolution of the   
    main part of a PN across the {Hertzsprung-Russell diagram (HRD)}.
   
   The different properties and evolutionary histories of shell and rim have also been found by 
   \citet{mellema.94, mellema.95} in his somewhat parameterised hydrodynamical studies on the 
   formation and evolution of PNe. Consequently, he coined the term \hbox{``I-shock''} for the
   leading shock of the shell in order to indicate that it is set up by ionisation, and the 
   term ``W-shock'' for the shock preceding the rim because the latter is the result of 
   wind-wind interactions.  Unfortunately, both terms could not find acceptance in the literature.  
           
   Because of the different expansion behaviour of rim and shell, their leading shocks are
   expected to have different properties and thus correction factors as well.
   In \citetalias{SBJ.18} the correction factors were carefully evaluated from the 
   hydrodynamical models.  As expected, both the inner bright rim and the outer shell demand
   different correction factors, reflecting the different driving mechanisms of rim and shell.  
   More details can be found in \citet{schoenetal.14} and  \citetalias{SBJ.18}. 
   
   In short, if $ \dot{R}_{\rm shell} $ is the propagation speed of the shell's shock front
   and $ V_{\rm post} $ the corresponding post-shock flow velocity\footnote
   {The determination of the flow velocity immediately behind the leading shock of the shell  
    by means of high-resolution spectrograms is discussed in detail in \citet{corradi.07}.
    }, 
   the relation
\begin{equation}
\label{eq:Fshell}   
   \dot{R}_{\rm shell} = F_{\rm shell}\times V_{\rm post} = (1.25 \pm 0.05)\times V_{\rm post} 
\end{equation}  
   holds,   
   virtually independent of the evolutionary state as long as the nebula is optically
   thin.  In fact, \citet[][Fig.~9 therein]{jacobetal.13} showed that $ F_{\rm shell} $ varies
   between about 1.2 and 1.4, depending on the parameters of the hydrodynamical sequences.
   The value chosen in \citetalias{SBJ.18} and used here is typical for nebula models around
   nuclei with masses below  0.6~\Msun.   
   If instead of the post-shock velocity only a ``bulk'' expansion velocity of the
   shell is available, the value of $ F_{\rm shell} $ is accordingly higher, namely about 1.5.
   
   The situation is different for the rim because (i) one can only measure the rim's
   bulk expansion, defined by half the peak separation of the split rim line emission,
   and (ii) the rim's expansion is strongly accelerating with time.  We write
\begin{equation}
\label{eq:Frim}   
\dot{R}_{\rm rim} = F_{\rm rim}(V_{\rm rim})\times V_{\rm rim} , 
\end{equation}  
   where $ F_{\rm rim} $ is a decreasing function of the rim velocity: $ F_{\rm rim} \simeq\! 3 $
   for  $ V_{\rm rim} = 5 $~\kms\ and $ F_{\rm rim} \simeq\! 1.5 $ for $ V_{\rm rim} = 20 $~\kms\
   (Fig.~5 in \citetalias{SBJ.18}). \changed{These numbers refer to measurements of \oiii\
   lines and may change if lines of other ions are used.}  
   We remark further that all studies prior to \citetalias{SBJ.18} used uncorrected rim 
   expansions only, leading to considerable underestimations of the distances by at least 50\%.  
   
   We repeat here that the correction factors to be applied to Eq.~(\ref{eq:Dexp}) must always  be
   larger than unity according to the physics of shock propagation \citep[e.g.][]{mellema.04}, 
   so that disregarding them will lead to a systematic underestimation of distances. 
   Support for the reliability of the correction factors as introduced and discussed here comes  
   from the fact that in the three cases where both rim and shell proper motions are available 
   the rim and shell distances agree within the errors \citepalias[Fig.~7 in][]{SBJ.18}.

\section{Expansion versus \g\ DR2 distances}
\label{sec:sample.comp}

\begin{table*}
\caption{\label{tab:sample}
 \changed{Our sample of 11 PNe with known expansion distances and existing \g\ DR2 trigonometric
          distances.   
         The table contains 
         the list of objects (Col.~1), the \g\ DR2 parallaxes (Col.~2), the corresponding
         distances (Col.~3), the expansion distances finally adopted by 
         \citetalias[][]{SBJ.18} (see Table~7 therein) (Col.~4), the uncorrected expansion 
         distances of the shells (Col.~5), the shell post-shock velocities (Col.~6), 
         the empirical shell correction factors (Col.~7), the uncorrected expansion velocities 
         of the rims (Col.~8), the rim (bulk) expansion velocities (Col.~9), and the empirical
         rim correction factors (Col.~10).  }      
         }
\center
\tabcolsep = 3.5pt
\begin{tabular}{l @{\hspace*{9mm}} c c c @{\hspace*{9mm}} c c c @{\hspace*{9mm}} c c c}
\hline\hline
\noalign{\smallskip}
   Object         &    DR2 Parallax          &  $ D_{\rm DR2} $          &  $ D_{\rm exp}  $ 
                  &$ D_{\rm exp,\, shell}^0 $& $ V_{\rm post} $          & $ F_{\rm shell} $
                  &  $ D_{\rm exp,\, rim}^0 $& $ V_{\rm rim} $  & $ F_{\rm rim} $ \\[2.5pt]
   
                  &     [mas]                &      [kpc]                     
                  &     [kpc]                &      [kpc]                  
                  &     [\kms]       &       &      [kpc] &   [\kms]     &    \\[1.5pt]
                  
\quad  (1)        &       (2)                &       (3)   &  (4)       
                  &       (5)                &       (6)   &  (7)  &  (8) & (9) & (10) \\[1.5pt]           
\hline\noalign{\medskip}   

\object{IC 418}   & $ 0.6453\pm 0.0541 $ & $ 1.550\,^{+0.141}_{-0.119} $ & $ 1.15\pm 0.20 $
                  & $ 0.88\pm 0.12 $     &  $ 22\pm 2 $ & $ 1.53\,^{+0.25}_{-0.18} $  
                  &   --    &    --        &     --                  \\ [3pt]
 
\object{IC 2448}  & $ 0.2883\pm 0.0454 $ & $ 3.469\,^{+0.647}_{-0.472} $ & $ 2.00\pm 0.30 $
                  & $ 1.20\pm 0.45 $     &  $ 35\pm 2 $  & $ 2.89\,^{+1,20}_{-1,15} $       
                  & $ 1.43\pm 0.14 $ & $ 18\pm 1 $  &  $ 2.42\,^{+0.50}_{-0.40} $   \\[3pt]

\object{IC 4593}  & $ 0.3803\pm 0.0794 $ & $ 2.630\,^{+0.693}_{-0.472} $ & $ 3.00\pm 1.20 $
                  & $ 2.40\pm 0.95 $     & $ 24\pm 2 $   & $ 1.10\,^{+0.52}_{-0.47} $ 
                  &   --    &    --      &     --                  \\[3pt]
 
\object{NGC 3132} & $ 1.1566\pm 0.0504 $ & $ 0.865\,^{+0.039}_{-0.036} $ & $ 1.25\pm 0.30 $
                  &       -- &   -- & -- 
                  & $ 0.89\pm 0.16 $ & $ 21\pm 2 $ & $ 0.97\,^{+0.18}_{-0.18} $  \\[3pt]

\object{NGC 3242} & $ 0.6819\pm 0.0884 $ & $ 1.466\,^{+0.219}_{-0.168} $ & $ 1.15\pm 0.15 $
                  &     --   &   -- & --
                  & $ 0.73\pm 0.09 $ & $ 17\pm 1 $ & $ 2.00\,^{+0.35}_{-0.33} $   \\[3pt]

\object{NGC 5882} & $ 0.5071\pm 0.0666 $ & $ 1.997\,^{+0.307}_{-0.254} $ & $ 1.70\pm 0.30 $
                  &     --   &   -- & --
                  & $ 1.21\pm 0.19 $ & $ 22\pm 1 $ & $ 1.65\,^{+0.36}_{-0.33} $    \\[3pt]

\object{NGC 6543} & $ 0.6152\pm 0.0709 $ & $ 1.625\,^{+0.212}_{-0.167} $ & $ 1.86\pm 0.15 $
                  &     --   &   -- & --
                  & $ 1.24\pm 0.09 $ & $ 19\pm 2 $ & $ 1.30\,^{+0.19}_{-0.17} $      \\[3pt]

\object{NGC 6826} & $ 0.6348\pm 0.0475 $ & $ 1.575\,^{+0.128}_{-0.103} $  & $ 1.55\pm 0.20 $   
                  & $ 1.20\pm 0.22 $     & $ 33\pm 2 $  & $ 1.31\,^{+0.25}_{-0.24} $ 
                  & $ 0.85\pm 0.10 $  & $ 8.5\pm 1~ $    & $ 1.86\,^{+0.27}_{-0.25} $  \\[3pt]

\object{NGC 6891} & $ 0.4073\pm 0.0512 $ & $ 2.455\,^{+0.353}_{-0.274} $ & $ 1.45\pm 0.45 $
                  &     --   &   -- & --
                  & $ 0.66\pm 0.16 $ & $ ~~7\pm 1 $  & $ 3.73\,^{+1.05}_{-0.99} $     \\[3pt]
 
\object{NGC 7009} & $ 0.8665\pm 0.1161 $ & $ 1.154\,^{+0.177}_{-0.136} $ & $ 1.50\pm 0.35 $
                  &     --   &   -- & --
                  & $ 1.00\pm 0.26 $ & $ 18\pm 1 $ & $ 1.15\,^{+0.35}_{-0.33} $      \\[3pt]

\object{NGC 7662} & $ 0.5054\pm 0.0747 $ & $ 1.977\,^{+0.345}_{-0.253} $ & $ 1.90\pm 0.30 $
                  & $ 1.32\pm 0.30 $     & $ 35\pm 3 $ & $ 1.50\,^{+0.43}_{-0.39} $
                  & $ 1.55\pm 0.13 $ & $ 26\pm 1 $ & $ 1.28\,^{+0.28}_{-0.23} $      \\[3pt]
  
\hline
\end{tabular}
\tablefoot{
  \changed{The uncorrected expansion distances for shell and rim in Cols.~5 and 8 
           follow from the data in Tables~4 and 5 of \citetalias[][]{SBJ.18}.  The velocities
           in Cols.~6 and 9 are in general mean values derived from the \nii~6548/6583~\AA\ and 
           \oiii~4959/5007~\AA\ lines as listed in \citet{schoenetal.14}. For optically thin
           nebulae, velocities derived from both ions are very similar (cf. Fig.~8 in
           \citealt{schoenetal.14}).}
          }
\end{table*}

\begin{figure}
\vskip -1mm
\includegraphics[trim= 1.7cm 0.2cm 0.5cm 1.7cm, clip, width=0.99 \linewidth]
                {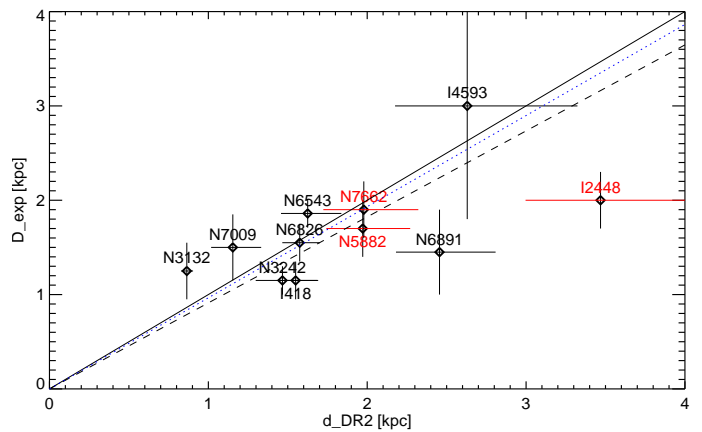}              
\caption{\label{fig:exp.gaia}
         Expansion distances, $ D_{\rm exp} $, of PNe versus the tri\-gon\-omet\-ric  
         \g\ $ D_{\rm DR2} $ distances (Cols.~3 and 4 of Table~\ref{tab:sample}).  
         The dashed line is the error weighted linear regression 
         forced to go through the origin (Eq.~\ref{eq:regress1}).   
 \changed{The dotted line represents the corresponding linear regression for the eight objects
          with ${ D_{\rm DR2} < 2 }$~kpc only (Eq.~\ref{eq:regress2}).} 
         The 1:1 relation is given for comparison as well (solid).   
         For the red objects (IC~2448, NGC~5882, NGC~7662), discrepancies between the terrestrial 
         and \g\ photometry of the central star exist (see text for details).
         }
\end{figure}

    We found 11 PNe in the \g\ DR2 catalogue that are in common with objects for which 
    \citetalias{SBJ.18} have derived expansion parallaxes.
    In general, the accuracy of the \g\ parallaxes is comparable to or even
    higher than for the expansion parallaxes.  Except for IC 4593, the errors of the \g\
    parallaxes are well below about 15\%, and in two cases, NGC~3132 and NGC 6826, even below 10\%. 
    In three cases (red in Fig.~\ref{fig:exp.gaia}), IC 2448, NGC 5882, and NGC 7662, 
    the \g\ photometry of the central star is not consistent with the 
    respective ground-based photometry. 
    However, we verified by the \g\ software services provided by Astromisches 
    Rechen-Institut, University of Heidelberg, that \g\ was locked onto the central star also
    in these cases.    Furthermore, the relative errors of \g`s \texttt{phot\_g\_mean\_flux} are
    very small, comparable to the other objects.  NGC 7009 has the highest flux error of the whole
    sample, namely 0.4\%.  Recalling that photometry of central stars from the ground may 
    sometimes be difficult because of a high background, we do not see any reason to exclude
    IC 2448, NGC 5882, and NGC 7662 from the \g\ comparison sample. 
           
    All the relevant data on expansion and trigonometric \g\ distances of our 11 sample objects 
    are given in Table~\ref{tab:sample}, supplemented by the uncorrected expansion distances as
    they follow separately for shell and rim (Cols. 5 and 8) 
    \changed{together with the corresponding (Doppler) expansion velocities 
    (Cols.~6 and 9) and correction factors (Cols.~7 and 10).}
    
    A comparison between \changed{the expansion and the \g\ distance sets} is displayed in
    Fig.~\ref{fig:exp.gaia}.
    Except for the two outliers IC~2448 and NGC~6891 which deserve further
    discussion, \changed{the data points follow closely a 1:1 relation.}  
    The linear regression \changed{considering all 11 objects} and
    forced to go through the origin is 
\begin{equation}
\label{eq:regress1}
     D_{\rm exp} = (0.912 \pm 0.059)\times D_{\rm DR2} ,  
\end{equation}  
    where the error bars in both directions were properly taken into account in the 
    determination of the best fitting slope and its error.
   \changed{Considering only the eight objects with \g\ DR2 distances below 2~kpc, we obtain} 
\begin{equation}
\label{eq:regress2}
     D_{\rm exp} = (0.966 \pm 0.066)\times D_{\rm DR2} .  
\end{equation}   
%

\subsection{Comments on individual objects}
\label{subsec:individual}

   Here we compare the new \g\ DR2 distances with other individual distances, if available,
   notably with those based on detailed stellar spectroscopy by \citet{MKH.92}, \citet{PHM.04},
   and \citet{KUP.06}, and/or with distances found from a modelling of the entire star-nebula 
   system.    Such a comparison is important because PNe with spectroscopically derived 
   distances have frequently been used as highly weighted calibrators for statistical distance
   scales in the past (see e.g. \citealt{frew.16}). 
   For the sake of brevity, we considered results from the more recent literature only.

\paragraph{IC 418}
\label{par:ic418}

   This object is the least evolved of our sample and on the verge of becoming 
   optically thin.
\changed{The visible PN is still surrounded by a huge photo-dissociation region (cf.   
   \citealt{gomez.18}).}
   We face the difficulty of choosing the proper combination of flow velocity and
   correction factor for the shell. 
   The two existing expansion parallax measurements of \citet{guzetal.09}   
   and \citetalias{SBJ.18} yielded distances of $ 1.30\pm 0.40 $~kpc and $ 1.15\pm 0.20 $~kpc,
   respectively, somewhat lower than the trigonometric distance found by \g, 
   $ 1.55\,^{+0.141}_{-0.119} $~kpc.  The higher trigonometric distance makes the nucleus
   of IC~418, with $ 11200\pm 2900 $~\Lsun, the most luminous (and also most massive, 
   $ \simeq\! 0.66 $~\Msun, \citealt{miller.16}) of  our sample,     
   comparable to the most luminous object in the Milky Way bulge sample of 
   \citet{hultzschetal.07}, which has $ \simeq\!10\,200 $~\Lsun.

   The \g\ DR2 distance to IC 418 agrees perfectly with the gravity distance derived by
   \citet{MKH.92}, 1.5~kpc, but with a somewhat lower stellar luminosity of 
   ${ \simeq\!9200 }$~\Lsun.   The photospheric analysis
   of \citet{MG.09} yielded a higher stellar gravity and consequently only a distance of 1.26~kpc
   and ${ \simeq\! 7600 }$~\Lsun\  for the central star.  
   Upscaling to the \g\ distance yields a stellar luminosity of ${ \simeq\! 11\,500 }$~\Lsun. 
   The distances of \citet{PHM.04} and \citet{KUP.06} are obviously too high: 2.0 and 2.6~kpc,
   respectively.  With these distances, the central-star luminosity gets extremely high,  
   ${ \simeq\! 16\,000 }$~\Lsun.

   Recently, \citet{dopetal.17} presented a detailed photoionisation model of the IC 418 system
   and came up with a distance of $ 1.0\pm 0.1 $~pc only.  This rather low value is certainly not
   compatible with \g's trigonometric distance.

\paragraph{IC 2248 and NGC 6891}
\label{par:ic2448.ngc6891}

   For these two objects, the trigonometric \g\ DR2 parallaxes are substantially lower than
   our expansion parallaxes, namely by a factor of about 1.7 in both cases.   
   We do not have a reasonable explanation for this discrepancy because 
   both objects have a rather regular elliptical shape and are thus  well suited to expansion 
   measurements. We note that our measured proper motion of NGC~2448's rim  agrees well
   with the older results of \citet{palenetal.02}, which are based on a shorter time span.
   The only possibility left to enlarge the expansion distance for both objects is to increase 
   the rim correction factors $ F_{\rm rim} $,
   namely from 1.5 to ${ \simeq\!2.4 }$ (IC~2448) and from 2.3 to ${ \simeq\!3.7 }$ (NGC~6891), 
   in contradiction with the predictions from our hydrodynamical models 
   \changed{(cf. Fig~\ref{fig:corr} below)}.  
   
   NGC 6891 is a twin of NGC 6826, and for the latter expansion and trigonometric
   distances agree very well (see below). The rim expansion of NGC 6891, with ${ 7\pm1 }$~\kms, 
   is the lowest of the whole sample.  At such a very slow expansion of the rim, $ F_{\rm rim} $ 
   is a steep function of $ V_{\rm rim} $ \changed{(Fig.~\ref{fig:corr})}.
   Given the uncertainty of $ V_{\rm rim} $, $ F_{\rm rim} $ values of 3 or even 
   higher are not unreasonable, which would then suffice for reaching a fair agreement with 
   the \g\ measurement. 
   
   This procedure will not work for IC 2448 because it is a rather evolved PN with a high value 
   of rim expansion (${ V_{\rm rim} = 18\pm 1 }$~\kms) and a correspondingly moderate value of 
   ${ F_{\rm rim} = 1.5 }$ (cf. Fig.~\ref{fig:corr}).
   Additionally, the distance of IC~2448 from the shell expansion is even lower but still
   consistent within the errors with the distance based on the rim expansion 
   (see Tables~4 and 5 in \citetalias{SBJ.18}).   
    A shell correction factor of about 2.9 (instead of 1.25) would be necessary in order to get
    agreement with the \g\ distance (cf. Table~\ref{tab:sample}).
    
   \citet{MKH.92} provided distances for both objects, 3.3~kpc (NGC 2448) and 3.0~kpc (NGC 6891),
   which are fully consistent with the respective \g\ trigonometric distances of 
   $ 3.469\,^{+0.647}_{-0.472} $~kpc (IC~2448) and $ 2.455\,^{+0.353}_{-0.274} $~kpc (NGC~6891).        
   Support for the higher \g\ distances comes from the fact that  
   the expansion distance results in very low stellar luminosities for both objects, 
   namely just a bit above 2000~\Lsun\ (see Table~8 in \citetalias{SBJ.18}).
   With the \g\ trigonometric distances, 
   more reasonable luminosities follow:  $ 6800\pm 1500 $~\Lsun\ for IC~2448 and 
   $6200\pm 1100 $~\Lsun\ for NGC~6891.

\paragraph{IC 4593}
\label{par:ic4593}

   The \g\ DR2 result confirms our high expansion distance, though the error bars of the latter
   are the largest of the whole sample.  
   The spectroscopic distances of \citet{MKH.92} (3.0~kpc) and 
   \citet{KUP.06} (3.3~kpc) are still consistent within the errors, while the distance of 
   \citet{PHM.04} (3.6~kpc) is too high.

\paragraph{NGC 3132}
\label{par:ngc3132} 

   This nebula has a far evolved low-luminosity nucleus, namely a hot white dwarf, and is thus
   optically thick \citepalias[cf.][]{SBJ.18}.  Our expansion distance is higher than the \g\
   distance, but here we have again the problem of a proper combination of flow and shock
   expansion velocities in \changed{recombined optically thick nebulae without a double-shell
   structure.} It appears that our chosen flow velocity or correction factor is somewhat too high. 
   \changed{We note that an empirical rim correction factor of about unity as found by us 
   (Col.~10 in Table~\ref{tab:sample}) is not in conflict with hydrodynamical PN models 
   in their final recombination/reionisation phase
   (see top panels of Fig.~A.1 in \citetalias{SBJ.18}).}  
   We also remark that NGC~3132 has the lowest distance of all the sample objects from
   Table~\ref{tab:sample} and thus also the smallest parallax error.

\paragraph{NGC 3242}
\label{par:ngc3242}   

   Already \citetalias{SBJ.18} gave a higher distance to NGC~3242 than found in earlier expansion
   works and discussed the reasons.  \citet{gomez.15} analysed the same HST images
   as were used by \citetalias{SBJ.18}, but came up with a distance of 
   $ 0.66\pm 0.10 $~kpc only.
   The reason is the neglect of the correction factor, which is about 1.6 in this particular case.
   Considering this factor, a distance of $ 1.06\pm 0.16 $~kpc follows, virtually the same as the
   \citetalias{SBJ.18} value of $ 1.15\pm 0.15 $~kpc, as one would expect 
   (compare Cols. 4 and 6 in Table~\ref{tab:sample}).  
   The \g\ value, $ 1.466\,^{+0.219}_{-0.168} $, is higher, but the errors just overlap.

   For this object, the spectroscopic gravity distances are all rather close to the expansion and 
   \g\ values:  
   1.7~kpc \citep{MKH.92, KUP.06} and 1.1~kpc \citep{PHM.04}.  With a distance to NCG~3242 of
   1.47~kpc, the luminosity of its nucleus increases from $\simeq$3000~\Lsun\ (1.15~kpc, Table~8 
   in \citetalias{SBJ.18}) to $ \simeq $\,5400~\Lsun, a very reasonable value.  
   
   In this context, we note that \citet{BK.18} published recently a photoionisation study 
   of the entire NGC~3242 system.  Using the wrong distance of 0.66~kpc of
   \citet{gomez.15}, they found for the central star a luminosity of ${ \simeq\!5500 }$~\Lsun,
   which is obviously an unrealistic combination of distance and luminosity:  
   scaling formally this luminosity value to the \g\ trigonometric distance of
   1.47~kpc, a completely unreasonable central-star luminosity of about 27\,000~\Lsun\ follows.

\paragraph{NGC 5882}
\label{par:ngc5882}   
 
   The detailed photoionisation modelling of \citet{milleretal.19} yielded a distance of
   $ 1.81\,^{+0.60}_{-0.82} $~kpc and a stellar luminosity of $ 2820\,^{+1650}_{-780} $~\Lsun.   
   This distance is well in between the expansion
   ($ 1.7\pm 0.30 $~kpc) and trigonometric \g\ ($ 1.997\,^{+0.307}_{-0.254} $~kpc) values.
   Using the \g\ parallax, the central-star luminosity increases somewhat, namely to 
   $ \simeq$3400~\Lsun.

\paragraph{NGC 6543}
\label{par:ngc6543} 

   The shell of NGC~6543 appears to be rather complex, but the rim is more regular and allowed
   a reliable expansion distance to be determined, $ 1.86\pm 0.15 $~kpc \citepalias{SBJ.18}, 
   which is only slightly higher than the \g\ trigonometric distance of 
   $ 1.625\,^{+0.212}_{-0.167} $~kpc.   The modelling of star and nebula by \citet{georgiev.08}
   yielded a distance of $ 1.8 $~kpc, but unfortunately the authors discarded this 
   value and replaced it by \citet{reed.99}'s expansion distance of 1~kpc and rescaled their 
   model accordingly.  \changed{Using the \g\ distance instead, a reasonable luminosity of
   4200~\Lsun\ follows.} 
   
   The expansion distance of \citet{reed.99} is, of course, too low because the authors did not 
   correct for the difference between pattern and matter velocities.  Applying the rim correction
   factor used by \citetalias{SBJ.18}, namely 1.5, \citeauthor{reed.99}'s expansion distance 
   fully agrees with the \g\ value.

\paragraph{NGC 6826}
\label{par:ngc6826}
  
   This is a very interesting object because \citet{PHM.04} found extreme values for
   central-star mass and luminosity:  1.4~\Msun\ and 15800~\Lsun\ in combination with a 
   distance of 3.2~kpc.  \g\ DR2 fully confirms the \citetalias{SBJ.18} expansion parallax in  
   that the distance to NGC 6826 is only half as much, namely $ 1.575\,^{+0.128}_{-0.103} $
   with a low error of only about 7\%.  The stellar luminosity would then be only about 
   3900~\Lsun, more in line with the mean ``plateau'' luminosity of the other sample stars of 
   ${ \simeq\! 5000 }$~\Lsun\ \citepalias{SBJ.18}.
   While \citet{KUP.06} estimated also a similarly high distance of 2.5~kpc, the gravity distance
   derived by \citet{MKH.92}, 1.8~kpc, is rather close to the \g\  value.
    
   There are several studies of the whole system available from the literature.  The most recent
   one is that of \citet[][Table~2 therein]{BK.18}  who found a distance of 
   ${ 2.1\pm 0.5 }$ and 6000~\Lsun. This distance is just compatible with the \g\ value within 
   the uncertainties.  We note, however, that their nebular modelling resulted in a stellar
   effective temperature of 65\,000~K, much too high for the spectral appearance of the star, 
   which suggests an effective temperature between 45\,000 and 50\,000~K only 
   (see e.g. \citealt{MKH.92, PHM.04, KUP.06}).  
 
   The earlier study of nebula plus star by \citet{fierro.11} came to a completely different
   result: ${ 0.8\pm 0.2 }$~kpc and 6000~\Lsun\ with ${ \Teff = 45\,000 }$~K.  
   Still other parameters were found by \citet{SP.08}.  From a Cloudy photoionisation model,
   the authors derived a best-fit stellar model with distance 1.40~kpc, 1640~\Lsun, and with
   ${ \Teff = 47\,500 }$~K.  \citealt{SP.08} presented arguments ``which  
   strongly rule out a higher luminosity than given by us''.
    
   It appears to us difficult to reconcile the results of these three studies.   Scaling the
   different luminosities found for the central star of NGC~6826 to the \g\ distance, 
   we end up with 3400~\Lsun\ \citep{BK.18}, 23400~\Lsun\ \citep{fierro.11}, 
   and 2090~\Lsun\ \citep{SP.08}.  On the other hand, the downscaling of the gravity distances
   to the \g\ distance leads to more realistic {but still diverse} luminosities for 
   the central star of  NGC~6826:  
   7200~\Lsun\ \citep{MKH.92}, 3900~\Lsun\ \citep{PHM.04}, and 4800~\Lsun\ \citep{KUP.06}.

\paragraph{NGC 7009}
\label{par:ngc7009}

   Although this object has an irregular rim, the expansion distance derived from the
   proper motion of the semi-minor axis agrees within the errors with the \g\ DR2 value
   (see Table~\ref{tab:sample}).\footnote
   {The distance of NGC~7009 listed in Table~8 of \citetalias{SBJ.18} is incorrect. It should read
    $ 1.5\pm 0.35 $~kpc as in Table~7 therein and in our Table~\ref{tab:sample}.
    The stellar parameters are not affected by this typo.
    }
   With the new \g\ distance, the central-star luminosity is lower than found by 
   \citetalias{SBJ.18}, namely from 4750 to 2800~\Lsun. The gravity distance of \citet{MKH.92} 
   is nearly twice as high as the \g\ distance: 2.1~kpc.

\paragraph{NGC 7662}
\label{par:ngc7662}

    Here again both the expansion and the trigonometric \g\ distances agree very well 
    (Table~\ref{tab:sample}),  but they do not agree with the much lower distance of 
    ${ 1.19\pm 0.15 }$~kpc given by \citet{BK.18} by means of photoionisation modelling.  
    As in the case of NGC 3242, this combination of distance with the stellar luminosity 
    $ \simeq\!5300 $~\Lsun\ appears to be strange because increasing the distance to the \g\
    value would result in an excessively high luminosity of about 15\,700~\Lsun.
    In contrast, the modelling of \citet{milleretal.19} provided a distance consistent with
    both the expansion and \g\ distances:  $ 1.80\,^{+0.46}_{-0.32} $, together with a 
    central-star luminosity of $ 5750\,^{+2370}_{-1290} $~\Lsun.

\section{Empirical determination of the hydrodynamical correction factors $\vec F$}
\label{sec:empir.corr}      
   
   Figure~\ref{fig:exp.gaia} and the discussion of the sample objects have already demonstrated 
   that distances derived by just combining line-of-sight Doppler velocities
   with proper motions of suited shock fronts must be corrected by appropriately chosen  
   factors, \changed{which are usually} well above unity. Tables~4 and 5  in \citetalias{SBJ.18}
   contain separately the information on shell and rim expansions, the used corrections, 
   and the distances that follow thereby.   For convenience, we give in
   Cols.~5 and 6 of Table~\ref{tab:sample} also the uncorrected distances that follow
   directly from the \citetalias{SBJ.18} expansion measurements by using Eq.~(\ref{eq:Dexp}).   
   With known trigonometric distances from \g,
   this opens up the possibility to determine empirically the correction factors for both the shell
   and rim separately, including also their dependence on evolution.
   
\begin{figure}
\center
\includegraphics[trim= 0.3cm 0.4cm 0.6cm 0.4cm, width=1.44\linewidth, angle= -90, clip]{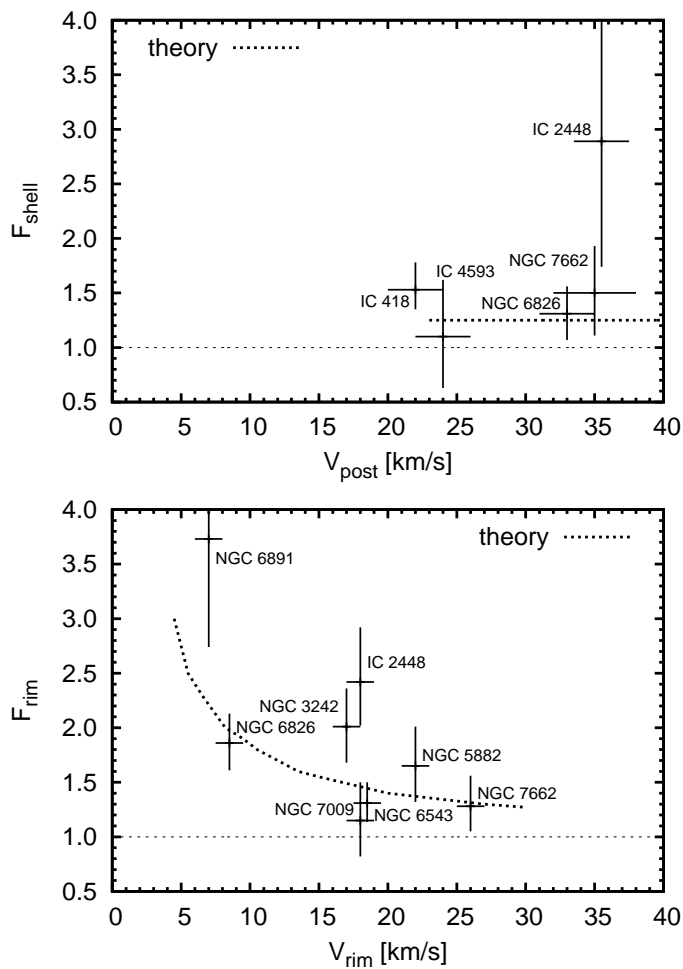}
\caption{\label{fig:corr}
         Correction factors $ F_{\rm shell} $ and $ F_{\rm rim} $, defined 
 \changed{according to Eq.~(\ref{eq:Fempirical}), }
         as function of the respective (line-of-sight) Doppler expansions:
         the shell's post-shock velocity $ V_{\rm post} $ (\emph{top panel}) and the rim's 
         bulk expansion velocity $ V_{\rm rim} $ (\emph{bottom panel}). 
         The crosses indicate the empirically determined $ F $ values for individual 
         objects, where \changed{their error bars are derived from the sums of the relative error
         squares of the trigonometric and expansion distances.} 
         The thick dotted lines are the theoretical expectations from hydrodynamical models,
 \changed{namely a constant value of $ 1.25 $ for the shell and an average from the model curves 
         in Fig.~5 of \citetalias{SBJ.18} for the rim.} 
         The weakly dotted horizontal lines highlight unity.
         Evolution is always from low to high velocities (see text for details).    
 \changed{The correction factor of NGC~3132 is not plotted here because this particular nebula
          is the most evolved of the whole sample and is in its recombination {or}
          reionisation phase, which is not covered by the hydrodynamical models employed for 
          the determination of the correction factors.}       
         }
\end{figure}
   
   The case is illustrated in Fig.~\ref{fig:corr} where we show \changed{(NGC~3132 excepted),}
   separately for shell and rim,
   the empirically determined correction factors \changed{from Table~\ref{tab:sample} (Cols.~7
   and 10).}      They are defined by the ratio between the \g\ trigonometric distance and the
   respective expansion distance without  considering any correction: 
\begin{equation}
\label{eq:Fempirical}   
    F_{\rm shell/rim} =  D_{\rm DR2} / D_{\rm exp,\ shell/rim}^0 \,.    
\end{equation}     
   We plotted the $ F $ factors against their respective Doppler velocities, $ V_{\rm post} $ 
   or $ V_{\rm rim} $, because the latter are a measure of the hydrodynamical state of a PN.  
   They are also a measure of the evolutionary progress since 
   it was shown empirically in \citet{schoenetal.14} that both velocities increase during 
   evolution across the HRD, albeit at different paces. 
      
   First of all, Fig.~\ref{fig:corr} confirms that the correction factors are always greater 
   than one for both the shell and rim expansion.   Their sizes and their run with evolution are,
   however, quite different.
\begin{description}

\item[Shell:]  $ F_{\rm shell} $ appears indeed to be independent of the shell's expansion, 
               which in turn also implies that the property of the shell's shock does not 
               really change while the central star crosses the HRD: the temperatures of the 
               four objects in the top panel of Fig.~\ref{fig:corr} span the range from about
               40\,000~K (IC~418 and IC~9543) to $\simeq$120\,000~K (NGC~7662). 
      \changed{With the exception of the ``outlier'' IC~2448,} the individual
               shell correction factors of the sample objects  are, within the errors, fully
               consistent with the prediction of our hydrodynamical models (thick dashed line at 
               $ F_{\rm shell} = 1.25 $).  

\item[Rim:]   The situation is different  for the rim (bottom panel).  Here, the correction factor
              $ F_{\rm rim} $ is predicted to decrease with increasing rim expansion.  This 
              trend is obviously confirmed  by the \g\ measurements.  
              The stellar temperature range is similar to that of the shell:  
              from about 50\,000~K (NGC~6826) to about 120\,000~K (NGC~7662).
      \changed{NGC~6891 has the lowest rim expansion and the biggest 
              empirical rim correction of all sample objects, although the uncertainty is
              very high.}                    
              
\end{description}   
   
     We especially comment on the two objects that are contained in both panels of 
     Fig.~\ref{fig:corr}: NGC~6826 and NGC~7662.   The former object is rather young with a 
     still slowly expanding rim, while NGC~7662 is far more evolved with its hot central star.
     Hence, the correction factors differ considerably (cf. Cols.~7 and 10 in
     Table~\ref{tab:sample}):  
     ${ F_{\rm shell} \simeq 1.3\pm 0.2 }$ and  ${ F_{\rm rim} \simeq 1.9\pm 0.3 }$ for NGC~6826, 
     in very close agreement with the prediction of our hydrodynamical models.  
     For the evolved object NGC~7662, both corrections are very similar: 
     ${ F_{\rm shell} \simeq 1.5\pm 0.4 }$ and  ${ F_{\rm rim} \simeq 1.3\pm 0.3 }$, 
     also consistent with the models.

\section{Summary and conclusion}
\label{sec:sum.con}

    We compared the trigonometric parallaxes of 11 PNe contained in the recent \g\ DR2 with 
    the corresponding expansion parallaxes that were based on multi-epoch HST images 
    in combination with Doppler expansion velocities and appropriate corrections derived from
    1-D hydrodynamical models.  From these 11 objects, only two objects, IC~2448 and NGC~6891,
    have discrepant distances for reasons we can only speculate about.  
    For the rest, a quite narrow 1:1 relation for
    the whole distance range from about 1 to 3~kpc exists.   This agreement could
    only be achieved by means of correction factors that take account of the hydrodynamics
    and of the fact that the expansion method combines pattern expansion (i.e. that of shock
    fronts) with matter velocities (gas flow velocities).
     
    Our main conclusion is that the \g\ data not only confirm the necessity of corrections 
    for the expansion parallaxes just as they have been derived in \citetalias{SBJ.18}, but also
    that they are different for rim and shell. This also proves that the 
    evolution of round {or} elliptical planetary nebulae is correctly described
    already by the 1-D hydrodynamical models of \citet{VMG.02} and 
    \citet{perinetal.98, perinetal.04}.   In particular, we also can conclude the following: 

\begin{itemize}
\item
    The well-known interacting winds paradigm for the formation and evolution of
    PNe cannot be applied to the shell, that is, to most of the nebula mass.  The shell's 
    formation is caused by photoionisation, and its following expansion is ruled by thermal
    pressure and the upstream density gradient.  The shell's shock does not ``know'' 
    the existence of a stellar wind.
    The latter only fills up nearly the entire cavity left by the expanding
    shell with shock-heated and thermalised wind matter (the hot bubble).  

\item   
    In contrast, the rim is the product of wind interactions   
    and is formed once the shell is established: 
    the increasing thermal pressure of the hot bubble forces the latter to expand and to compress
    the inner parts of the shell, 
    and the rim's further expansion is ruled by the bubble's
    pressure, which in turn depends on the evolution of the stellar wind luminosity.
\item   
   The \g\ DR2 results also confirm the findings of \citetalias{SBJ.18} that the 
   spectroscopically derived distances have the tendency to be overestimated.
   This fact needs further investigations and may also lead to an improvement of the 
   stellar atmosphere and/or wind modelling.   
\item
   Really disturbing is the fact that elaborate analyses of the combined system, nebula and 
   central star, may not always lead to parameters consistent with the values inferred from
   the \g\ distances.
\end{itemize}    
    
    Finally, we note that the (geometrically) thin rim is subject to dynamical instabilities, 
    but these are obviously not as strong as the \hbox{2-D} simulations of spherical model nebulae
    performed by \citet{TA.16} suggest.  Otherwise,  rims as observed would not exist, and the
    determination of reasonable expansion parallaxes from the rims would not have been possible.    
    It is hoped that future 3-D simulations with sufficient spatial resolution will improve 
    this issue.

\begin{acknowledgements}
    We are very grateful to Dr. Stefan Jordan, Astronomisches Rechen-Institut am Zentrum f\"ur
    Astronomie der Universit\"at Heidelberg, for his help in interpreting the
    relevant \g\ DR2 data.   \changed{We thank the referee for his or her very useful comments
    that helped us to improve the presentation of the paper.}
\end{acknowledgements}


\end{document}